\documentclass[fleqn,usenatbib]{mnras}

\usepackage{newtxtext,newtxmath}
\usepackage{mathrsfs}
\usepackage[normalem]{ulem}
\usepackage[T1]{fontenc}
\usepackage{ae,aecompl}
\usepackage{booktabs}

\usepackage{graphicx}	% Including figure files
\usepackage{amsmath}	% Advanced maths commands
\usepackage{float}

\usepackage{empheq}

\title[Molecular gas kinematics of FRB180924B host]{The molecular gas kinematics in the host galaxy of non-repeating FRB 180924B}

\author[T.-Y. Hsu et al.]{
Tzu-Yin Hsu,$^{1}$\thanks{E-mail: emma30407@gmail.com}
Tetsuya Hashimoto,$^{2}$
Bunyo Hatsukade,$^{3}$
Tomotsugu Goto,$^{4}$
Po-Ya Wang,$^{1}$
\newauthor
Chih Teng Ling,$^{4}$
Simon C.-C. Ho,$^{4,5}$
and Yuri Uno$^{2}$
\newauthor
\\
$^{1}$Department of Physics, National Tsing Hua University, 101, Section 2. Kuang-Fu Road, Hsinchu, 30013, Taiwan (R.O.C.)\\
$^{2}$Department of Physics, National Chung Hsing University, No. 145, Xingda Rd., South Dist., Taichung, 40227, Taiwan (R.O.C.)\\
$^{3}$Institute of Astronomy, Graduate School of Science, The University of Tokyo, 2-21-1 Osawa, Mitaka, Tokyo 181-0015, Japan\\
$^{4}$Institute of Astronomy, National Tsing Hua University, 101, Section 2. Kuang-Fu Road, Hsinchu, 30013, Taiwan (R.O.C.)\\
$^{5}$Research School of Astronomy and Astrophysics, The Australian National University, Canberra, ACT 2611, Australia\\
}

\date{Accepted 2022 December 06. Received 2022 December 02; in original form 2022 October 20}

\pubyear{2022}

\begin{document}
\label{firstpage}
\pagerange{\pageref{firstpage}--\pageref{lastpage}}
\maketitle
\begin{abstract} %250 currently 250 limit
Fast radio bursts (FRBs) are millisecond-duration transients with large dispersion measures. 
The origin of FRBs is still mysterious. 
One of the methods to comprehend FRB origin is to probe the physical environments of FRB host galaxies. 
Mapping molecular-gas kinematics in FRB host galaxies is critical because it results in star formation that is likely connected to the birth of FRB progenitors.
However, most previous works of FRB host galaxies have focused on its stellar component.
Therefore, we, for the first time, report the molecular gas kinematics in the host galaxy of the non-repeating FRB 180924B at $z= 0.3216$. 
Two velocity components of the CO (3-2) emission line are detected in its host galaxy with the Atacama Large Millimeter/submillimeter Array (ALMA): the peak of one component ($-155.40$ km s$^{-1}$) is near the centre of the host galaxy, and another ($-7.76$ km s$^{-1}$) is near the FRB position.
The CO (3-2) spectrum shows asymmetric profiles with A$_{\rm peak}$ $=2.03\pm 0.39$, where A$_{\rm peak}$ is the peak flux density ratio between the two velocity components. 
The CO (3-2) velocity map also indicates an asymmetric velocity gradient from $-180$ km s$^{-1}$ to 8 km s$^{-1}$.
These results indicate a disturbed kinetic structure of molecular gas in the host galaxy. 
Such disturbed kinetic structures are reported for repeating FRB host galaxies using HI emission lines in previous works.
Our finding indicates that non-repeating and repeating FRBs could commonly appear in disturbed kinetic environments, suggesting a possible link between the gas kinematics and FRB progenitors. 

%The property (e.g., center position, CO luminosity, molecular gas mass, center velocity, and dispersion velocity) of the two components in the CO spectrum is provided. 
%We compare the gas kinetic in the previous HI observations which indicate disturbing and merging HI gas structures for the host galaxy of repeating FRB 20180916B and the interacting galaxy of the host galaxy of repeating FRB181030A, and FRB200120E with increasing star forming rate. The CO emission spectrum of the FRB 180924B host galaxy shows asymmetric. 
%The velocity map shows a smooth gradient with a velocity offset. 
%Therefore, the disturbed kinetic state of molecular gas in the non-repeating FRB host galaxy is similar to the kinetic states of repeating FRB host galaxies which is also disturbed. We did not find a significant difference in the gas kinetic around repeating and non-repeating FRB.
%Therefore, the smooth rotation of molecular gas in the non-repeating FRB host galaxy is in contrast to the kinetic states of repeating FRB host galaxies. Our finding indicates different kinetic environments around different populations of FRBs, supporting a hypothesis that non-repeating and repeating FRBs originate from different progenitors.

\end{abstract}


%%%%%%%%%%%%%%%%%%%%%%%%%%%%%%%%%%%%%%%%%%%%%%%%%%

%%%%%%%%%%%%%%%%% BODY OF PAPER %%%%%%%%%%%%%%%%%%

\section{Introduction}
\label{introduction}
Fast radio bursts (FRBs) are millisecond-duration, energetic, and highly dispersed radio pulses \citep{Lorimer2007}. 
Most of them are regarded as extragalactic transients due to their large dispersion measures \citep[e.g.,][]{Thornton2013, Cordes2019} and localization to the host galaxies \citep[e.g.,][]{Tendulkar2017, Bannister2019, Nimmo2022}.
FRBs include two types, repeating FRBs and non-repeating FRBs. 
Whether the two populations of FRBs originate from similar progenitors is still highly debated \citep[e.g.,][]{Hashimoto2020, Pleunis2021}. 
So far, there are more than 800 FRBs detected, and numerous models have been proposed \citep[e.g.,][]{Margalit2018, Kashiyama2017, Lyutikov2016, Platts2019} to understand the progenitor of FRBs. 
However, the origin of FRBs remains a mystery.
One of the methods to comprehend the origin of FRBs is to probe the physical environments of their host galaxy. 
Most previous studies of FRB host galaxies are mainly focusing on its stellar components \citep[e.g.,][]{Bhandari2022, Mannings2021, Heintz2020}. 
The optical work of \citet{Bhandari2022}, which reported diverse properties of FRB host galaxies, suggests that there is no significant difference between the physical properties of host galaxies of repeating and non-repeating FRBs.
The mass-weighted stellar-mass cumulative distributions in \citet{Bhandari2022} indicate that FRB host galaxies are less massive than typical star-forming galaxies  with the p-value of P$_{\rm KS}=0.002$ for the Kolmogorov-Smirnov test.

HI-gas (molecular-gas) kinematics probed by radio (submillimeter) observations allows us to investigate variable physical processes of host galaxies (e.g., infalling gas, merger, and smooth rotation). 
The previous study of HI distribution in the host galaxy of repeating FRB 180926B \citep{Kaur2022} indicates the disturbed and merging kinetic structure of HI gas. 
\cite{Kaur2022} argues that the merging gas caused the burst of star formation in the outskirts of the host galaxy, which might be linked to the progenitor activity.
Asymmetric HI spectral shapes are reported for the host galaxies of repeating FRB 181030A, FRB 200120E, and FRB200428, indicating features of galaxy interactions \citep{Michalowski2021}.
The interactions found in these repeating FRB host galaxies indicate that there might be a connection between the birth of the FRB progenitors and recently enhanced star formation via interaction \citep{Michalowski2021}.

Since stars are born from molecular gases, they provide us with a more direct probe of kinetic features and a recent enhancement of star formation than HI gas. Therefore, mapping molecular-gas kinematics with CO observations in FRB host galaxies is critical, and the kinematics gives rise/decline to recent star formation that is likely connected to the origins of two FRB populations: non-repeating and repeating FRBs. 
Although understanding molecular-gas kinematics provides an insightful perspective, it has been poorly understood so far. To overcome this problem, we, for the first time, investigate molecular-gas kinematics in a host galaxy of a non-repeating FRB 180924B to understand the progenitor environments with the Atacama Large Millimeter/submillimeter Array (ALMA). 

\citet{Hatsukade2022} discussed the diverse molecular-gas properties in FRB host galaxies based on ALMA CO observations. The host galaxy of FRB 180924B is more gas-rich than the general star-forming galaxies \citep{Hatsukade2022}. They used a moderate velocity resolution of 50 km s$^{-1}$ to enhance CO detection. In this work, we use the same data as used by \citet{Hatsukade2022} with a better velocity resolution of 9 km s$^{-1}$ to focus on the kinematics of molecular gas. We report the first analysis of the kinetic structure of the molecular gas in the host galaxy of non-repeating FRB 180924B. The paper is organized as follows:
In section \ref{data analysis}, we present the ALMA data and data reduction. Section \ref{Kinetic stucture} demonstrates the kinetic structure in molecular gas. We discuss the possible environmental similarity in terms of gas kinematics between repeating FRBs and non-repeating FRBs in Section \ref{discussion}.
\begin{table*}

\centering
\caption{Physical properties of each peak}
\label{table1}
    \begin{tabular}{lccccccc}
        \hline
        Component   & RA                             & Dec                                                      & $V_{\rm sys}$   & $\sigma_{V}$ & S/N     & $L_{\rm CO}$                    & $M_{\rm gas}$                \\
                    &                                &                                              & (km s$^{-1}$)        & (km s$^{-1}$)      &         & (K km s$^{-1}$ pc$^{2}$)                    & ($M_{\odot})$                \\
        \hline
        Component 1 & $21 h 44m 25.2650 \pm 0.0034 s$ & $-40^{\circ} 54\arcmin 01.0496\arcsec \pm 0.0271\arcsec$ & $-155.40$ & $16.40$ & $ 9.60$ & $(9.35\pm0.97)\times10^{7}$ & $(7.31\pm0.76)\times10^{8}$  \\
        \hline
        Component 2 & $21 h 44m 25.2511 \pm 0.0037 s$ & $-40^{\circ} 54\arcmin 00.5665\arcsec \pm 0.0394\arcsec$ & $-7.76$   & $58.60$ & $8.76$  & $(1.65\pm0.18)\times10^{8}$ & $(1.28\pm0.14)\times10^{9}$ \\
        \hline
        
    \end{tabular}
    
\end{table*}
\begin{figure*}
    	\includegraphics[width=2\columnwidth]{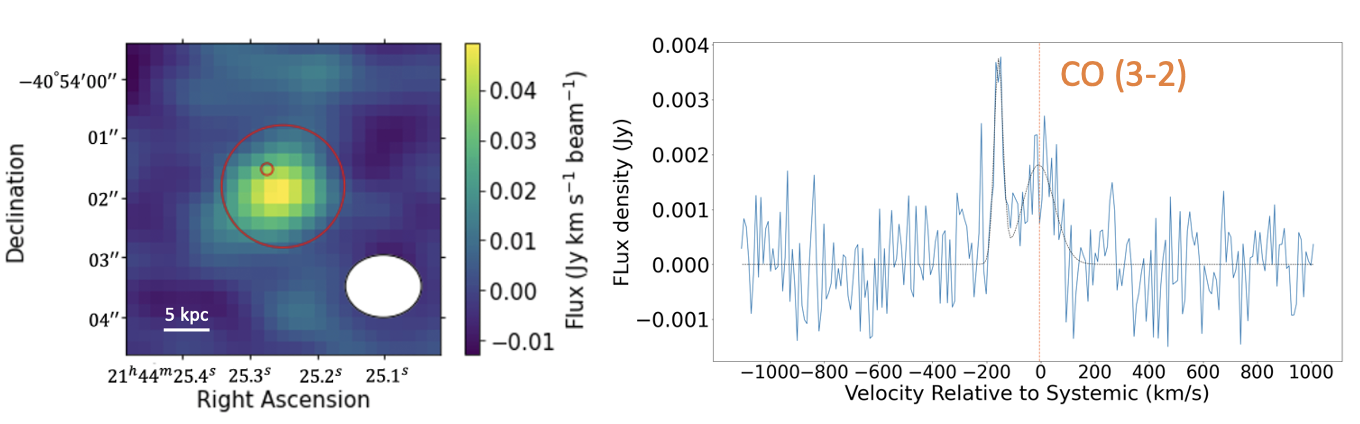}
    	\caption{(left) CO velocity-integrated intensity map of the host galaxy of FRB 180924B. The velocity range from $-180$ to 110 km s$^{-1}$ is integrated. The large red circle indicates the position of the host galaxy with a radius of twice its half-light radius \citep{Bhandari2020}. The small red circle indicates the FRB position there with an error as its radius \citep{Bannister2019}.  The beam size of ALMA data is shown by a white ellipse at the bottom right corner. (right) The spectrum of the CO emission line in the host galaxy of FRB 180924B. It is extracted from the aperture with a radius of 1.032 \arcsec (5 kpc) centred at the host galaxy (large red circle in the left figure). The blue line shows the observed data, the black line shows the double-Gaussian fit to the data, and the orange dashed line represents the systemic velocity at $z = 0.3214$.} 
    	
        \label{fig.1}
\end{figure*}        

%%%%%%%%%%%%%%%%% Data analysis %%%%%%%%%%%%%%%%%%

\section{Observation and Data analysis}
\label{data analysis}
%\subsection{ALMA data}
%\label{ALMA data}
The observation of FRB 180924B was conducted with Australia Square Kilometre Array Pathfinder (ASKAP) on 2018 September 24 at 16:23:12.6265 UTC \citep{Bannister2019}. 
The host is a massive spiral galaxy at $z=0.3214$ with a stellar mass of $M_{\rm \ast}=(1.32 \pm 5.1)\times 10^{9} M_\odot$ and the star formation rate (SFR) of $0.88 \pm 0.26 M_\odot$ yr$^{-1}$ \citep{Bannister2019, Heintz2020}. 
We utilized ALMA Science Archive data of Band 6 CO (3-2) observation of the FRB 180924B host galaxy with the project code 2019.01450.S\footnote[1]{\url{https://www.youtube.com/watch?v=qz92y4AtpDo}}. The host of FRB 180924B was observed using 44--46 antennas in array configuration with 15.0--311.7 m baseline length on November 26, 2019, for a total of 27 minutes on the source. The bandwidth of the correlator setup is 1.875 GHz, subsided into 240 channels.

The data reduction is conducted with Common Astronomy Software Application \citep[CASA;][]{McMullin2007}, and the calibration data is processed with the CASA pipeline version 5.6.1-8.
The data cube is obtained with the \texttt{CASA} task \texttt{tclean()} by setting the robust parameter of $2.0$ for Briggs weighting with a velocity resolution of $\sim$9 km s$^{-1}$. 
The observed frequency of the data cube is converted to the rest-frame frequency using the spectroscopic redshift of $z=0.3214$ \citep{Bannister2019}.
The Doppler velocity is calculated using the rest-frame frequency of the CO (3-2) emission line, i.e., 345.796 GHz.
The velocity-integrated intensity (moment 0) map and the velocity structure (moment 1) map are obtained by Cube Analysis and Rendering Tool for Astronomy (CARTA) \citep{Comrie2021}.
The synthesized beam of the two maps is $0 \arcsec.51 \times 0\arcsec.68$. 
%\section{Kinetic structure of molecular gas}
\section{Results}
\label{Kinetic stucture}
\subsection{CO emission line detection}
\label{CO emission line detection}
Figure \ref{fig.1} (left) shows the velocity-integrated intensity map with the integration range of $\sim -180$ km s$^{-1}$ to $\sim 110$ km s$^{-1}$. 
The CO (3-2) emission line is significantly detected at the position of the host galaxy within two-times half-light radius ($r_{\rm half}$): $r_{\rm half}=$ 0.516 arcsec $\sim$ 2.5 kpc \citep{Bhandari2020}. 
We utilize an aperture with a radius of 1.032 arcsec (5 kpc) centred at the host galaxy (RA, Dec)$=$ (21h44m25.25s, $-40^{\circ} 54\arcmin 00.81\arcsec$) \citep{Bhandari2020} to extract a CO spectrum of the host galaxy.
Figure \ref{fig.1} (right) shows the extracted CO spectrum.
In figure \ref{fig.1} (right), two velocity components are detected.
We fit double Gaussian functions to these two components in the spectrum. 
The best-fit double Gaussian functions show the systematic velocities ($V_{\rm sys}$) of $-155.40$ km s$^{-1}$ and $-7.76$ km s$^{-1}$ with velocity dispersions ($\sigma_{V}$) of 16 km s$^{-1}$ and 58 km s$^{-1}$, respectively.
\begin{figure}
    	\includegraphics[width=0.5\textwidth]{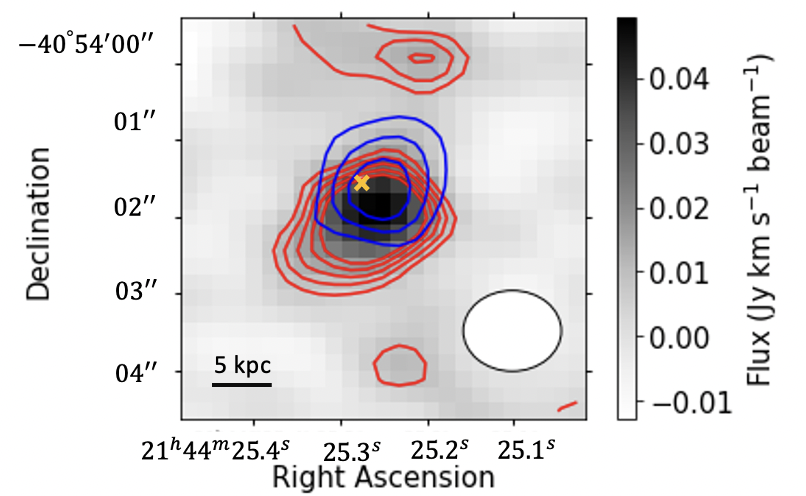}
    	\caption{The positional offset between the two CO (3-2) emission-line components are overplotted on the velocity-integrated intensity map (figure \ref{fig.1}) in grey scale.
      Blue and red contours indicate the flux level of the first  (centred at $-155.40$ km s$^{-1}$ in figure \ref{fig.1}) and second (centred at $-7.76$ km s$^{-1}$) velocity components, respectively. The centres of the contours are the position of two velocity components, which are present in table \ref{table1}.
     The contour starts from 2.5 $\sigma$ level with 1 $\sigma$ increment. We note that all the contours are not drawn for clarity. 
     The grey scale corresponds to the emission-line flux integrated over the velocity range from $-180$ km s$^{-1}$ to 110 km s$^{-1}$ that is extracted from the CO spectrum in figure \ref{fig.1}. 
     The yellow cross represents the position of FRB 180924B \citep{Bannister2019}.}
    	
        \label{fig.2}
\end{figure}  
\subsection{Positional offset between two peaks}
\label{Positional offset of two peaks}
Using the \texttt{CASA} task \texttt{imfit}, we obtain the central position of each peak of the two velocity components. The positional information and the physical properties of two velocity components are listed in table \ref{table1}. The distance between the two peaks is $\sim 0.\arcsec51$, which is more than 5 $\sigma$ apart with the positional uncertainty ($e$) of $\sim 0.\arcsec10$ that is derived from the following equation:
\begin{equation}
\label{positional uncertainty} 
e = \frac{a}{S/N},
\end{equation}
where $a$ is the major axis of the beam size in arc second. Both $a$ $\sim 0.\arcsec68$ and S/N $= 9.60$ and $8.76$ (table \ref{table1}) are obtained from the velocity-integrated intensity map in figure \ref{fig.2}. We calculate a quadrature sum of the positional errors of the two velocity components to derive $\sim 0.\arcsec10$ positional uncertainty.

Figure \ref{fig.2} shows the velocity-integrated intensity map with contours of different flux levels of two velocity components. The map shows multiple spatial components, where the blue contour represents the first component with $V_{\rm sys} = -155.40$ km s$^{-1}$ while the red contour represents the second component with $V_{\rm sys} = -7.76$ km s$^{-1}$. Each velocity component is integrated over $\pm$ 2 $\sigma_{V}$ from $V_{\rm sys}$.
The centre of the blue contour is relatively close to the FRB position, while the centre of the red contour is close to the host galaxy centre.

The CO luminosity and molecular gas mass are derived from the CO(3-2) emission-line flux measurement for each component. 
The CO Luminosity is calculated using equation (3) in \citet{Solomon2005} as follows:
\begin{equation}
\label{luminosity equ} %luminosity equ
L'_{\rm CO (3-2)}=3.25\times 10^{7} S_{\rm CO (3-2)}\Delta{\rm V} {\nu}_{\rm obs}^{-2} D_L^{2} (1+z)^{-3},
\end{equation}
where $L'_{\rm CO (3-2)}$ is measured in K km$^{-1}$ s$^{-1}$ pc$^{2}$, S$_{\rm CO (3-2)}\Delta$V is the velocity integrated flux in Jy km s$^{-1}$, ${\nu}_{\rm obs}$ is the observed frequency of the CO(3-2) emission line in GHz, and D$_L$ is the luminosity distance to the host galaxy in Mpc.
The Molecular gas mass is derived as follows:
\begin{equation}
\label{molecular gas equ}
M_{\rm gas}=\alpha_{\rm CO} L'_{\rm CO(1-0)}.
\end{equation}

The CO (1-0) luminosity ($L'_{\rm CO(1-0)}$) is derived by applying the CO luminosity ratio of $L'_{\rm CO(3-2)}/L'_{\rm CO(1-0)}=0.55$ for star-forming galaxies \citep{Lamperti2021,Hatsukade2022}. 
The conversion factor $\alpha_{\rm CO}$ is regarded as related to the gas-phase metallicity. The increasing  $\alpha_{\rm CO}$ indicates the decreasing gas-phase metallicity \citep{Bolatto2013_2}. We adopted $\alpha_{\rm CO}=4.3$ $M_\odot$ (K km$^{-1}$ s$^{-1}$  pc$^{2}$)$^{-1}$
since the gas-phase metallicity is $12 + \log ({\rm O/H})=8.93 \pm 0.02 $ \citep{Heintz2020}, which is comparable to the Milky way gas-phase metallicity.

%The adopted conversion factor $\alpha_{\rm CO}$ is the Milky-Way value of 4.3 $M_\odot$(K km$^{-1}$ s$^{-1}$)$^{-1}$ pc$^{2}$ \citep{Bolatto2013}.

The CO luminosity and molecular gas mass of each peak are demonstrated in Table \ref{table1}.
The total CO luminosity (molecular-gas mass) of the two velocity components is 
($2.59 \pm 0.20) \times 10^{8}$ K km s$^{-1}$ pc$^{2}$ (($2.01 \pm 0.16) \times 10^{9}$ M$_\odot$). This value is consistent with ($3.1 \pm 0.3) \times 10^{8}$ K km s$^{-1}$ pc$^{2}$(($2.4 \pm 0.2) \times 10^{9}$ M$_\odot$) reported by \citep{Hatsukade2022} within the errors. 

\subsection{Asymmetric CO spectrum}
\label{Asymmetric of CO spectrum}
The right panel of figure \ref{fig.1} shows the asymmetric profile of the CO(3-2) spectrum, which could indicate the disturbed kinetic structure of molecular gas. To test the asymmetry of the CO spectrum, we apply the diagnostic \citep{Reynolds2020} to define the spectrum asymmetry, deriving the ratio of the integrated fluxes between the left and right halve of the spectrum, A$_{\rm flux}$ \citep[equation 6 in][]{Reynolds2020} and the peak ratio of the two velocity components, A$_{\rm peak}$ \citep[equation 7][]{Reynolds2020}.
To derive A$_{\rm flux}$, the boundary between the left and right halve of the spectrum is defined by the mid-point of the full width at 20 \% of the maximum peak height as $V_{\rm mid}$. We also define the velocity that increases to the 20 \% of peak flux density in the first component as $V_{\rm low}$ and the velocity that decreases to the 20 \% of peak flux density in the second component as $V_{\rm high}$.
We integrate the best-fit Gaussian function from $V_{\rm low}$ to $V_{\rm mid}$ for the left halve and $V_{\rm mid}$ to $V_{\rm high}$ for the right halve. If A$_{\rm flux}$ or A$_{\rm peak}$ $<1$, we take the inverse values such that both A$_{\rm flux}$ and A$_{\rm peak}$ are higher than 1.

\begin{figure*}
    	\includegraphics[width=2\columnwidth]{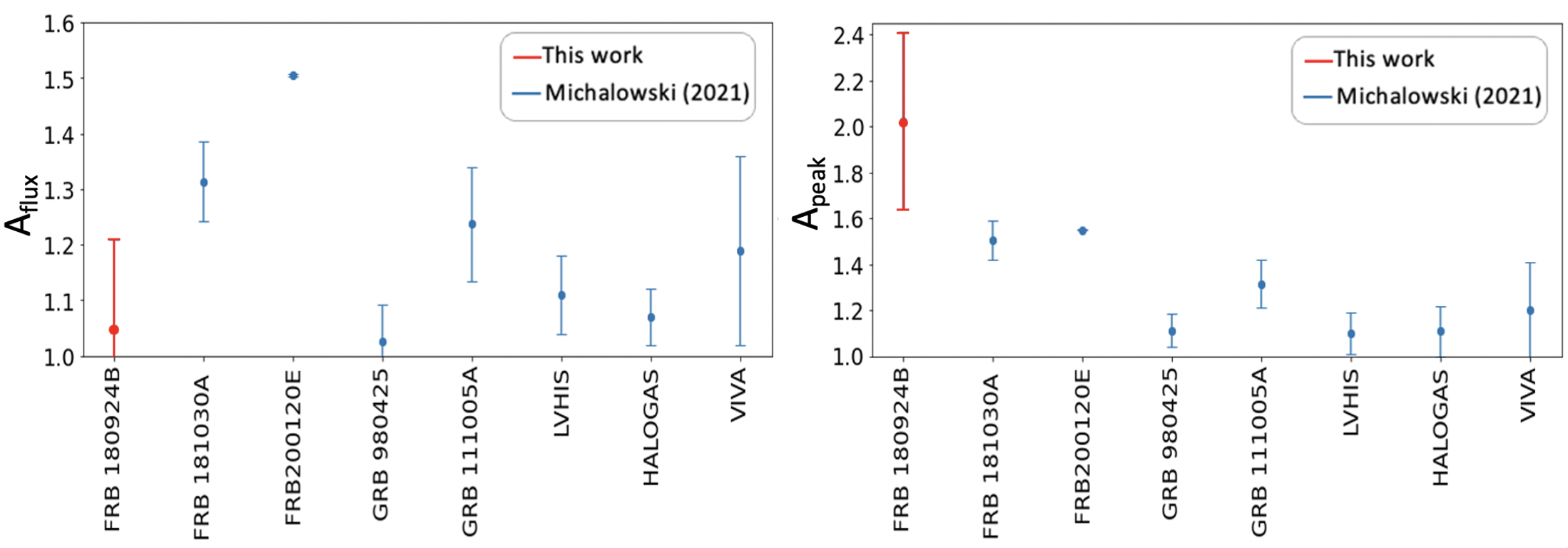}
    	\caption{Asymmetry diagnostics: (left) the ratio of the integrated fluxes between the left and right halve of the spectrum, A$_{\rm flux}$, with the boundary defined by the mid-point of the full width at 20 \% of the maximum peak height as $V_{\rm mid}$. The velocity that increases to the 20 \% of peak flux density in the first component is defined as $V_{\rm low}$ and the velocity that decreases to the 20 \% of peak flux density in the second component is defined as $V_{\rm high}$. We integrate the best-fit Gaussian function from $V_{\rm low}$ to $V_{\rm mid}$ for the left halve and $V_{\rm mid}$ to $V_{\rm high}$ for the right halve. (right) the ratio of the flux densities of the two peaks. The asymmetry parameter is compared with the FRB host galaxies, GRB host galaxies, and galaxy survey, i.e., LVHIS, HALOGAS, and VIVA  in \citep{Michalowski2021}. Compared with other galaxies, FRB180924B host galaxy shows high A$_{\rm peak}$.  If A$_{\rm peak}$ or A$_{\rm flux}$ $<1$, we take the inverse value.}
    	
        \label{fig.3}
\end{figure*}

\begin{figure}
    	\includegraphics[width=0.5\textwidth]{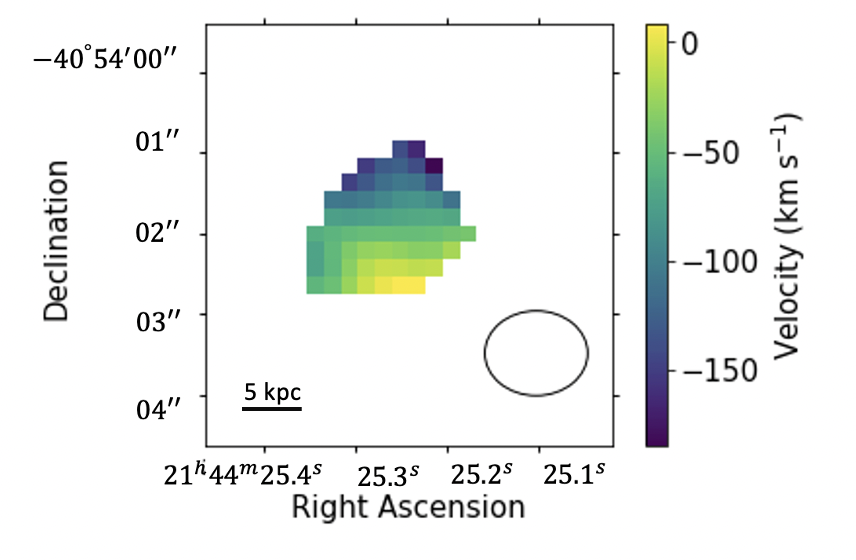}
    	\caption{
     Velocity map of the CO (3-2) emission of the FRB 180924B host. The region is chosen for the CO emission higher than 4 $\sigma$. The colours indicate molecular gas velocities at different positions. An ellipse shows the beam size of ALMA data at the bottom right corner.}
        \label{fig.4}
\end{figure}

 We compare our results with the asymmetry parameters of the HI emission lines in previous works, i.e., the host galaxy of FRB 181030A, FRB200120E, the host galaxy of long gamma-ray burst (GRB) GRB 980425 and GRB 111005A \citep{Michalowski2021}, the galaxies with stellar masses of $9<\log(M_{\rm star}/M_{\odot})<10$ in Local Volume HI Survey (LVHIS), Hydrogen Accretion in Local Galaxies Survey (HALOGAS), and VLA Imaging of Virgo in Atomic Gas (VIVA) \citep{Michalowski2021, Koribalski2018, Chung2009, Heald2011}. Figure \ref{fig.3} and table \ref{table3} shows the comparison of the asymmetry parameters. A$_{\rm flux}$ and A$_{\rm peak}$ of the FRB 180924B host are $1.05 \pm 0.16$ and $2.03 \pm 0.39$, respectively. The high value of A$_{\rm peak}$ indicates significant asymmetry in the CO spectrum and the disturbed kinetic structure of molecular gas.
Following \citet{Reynolds2020}, we define the spectrum as asymmetry with A$_{\rm flux}>1.05$, which is commonly used in the early studies \citep[e.g.,][]{Reynolds2020}. The comparison sample by \citet{Michalowski2021} also uses 1.05, allowing a fair comparison with previous studies.
A$_{\rm flux}$ of the host galaxy of FRB 180924B is on the boundary of the asymmetry, whereas A$_{\rm peak}$ indicates the clear asymmetric kinematics.
This might be because the kinetic information could be lost by the integration process \citep[equation 6 in][]{Reynolds2020}, making A$_{\rm flux}$ less sensitive to the spectral shape of FRB 180924B host.
Furthermore, figure \ref{fig.4} shows the velocity field of the region with CO (3-2) emission-line detection higher than $4 \sigma$. The velocity map indicates a smooth velocity gradient from $\sim -180$ km s $^{-1}$ to $\sim 8$ km s $^{-1}$. The missing positive velocity component indicates the asymmetric velocity gradient, suggesting a disturbed kinetic structure of molecular gas instead of a smooth and symmetric rotation. 

\begin{table}
\centering
\caption{Comparison of asymmetry parameters with different galaxies \citep{Michalowski2021}}
\label{table3}
    \begin{tabular}{lcc}
        \hline
         Host galaxy  & A$_{\rm flux}$ & A$_{\rm peak}$  \\         
        \hline
         FRB 1809024B & $1.05 \pm 0.16$   & $2.03 \pm 0.39$  \\   
        \hline   
         FRB 181030A  & $1.314 \pm 0.072$ & $1.505 \pm 0.084$  \\ 
        \hline   
         FRB 200120E  & $1.505 \pm 0.002$ & $1.551 \pm 0.004$  \\ 
        \hline   
         GRB 980425  & $1.027 \pm 0.066$ & $1.112 \pm 0.071$  \\ 
        \hline   
         GRB 111005A  & $1.237 \pm 0.102$ & $1.314 \pm 0.104$  \\ 
        \hline   
         LVHIS  & $1.110 \pm 0.070$ & $1.100 \pm 0.090$  \\
        \hline   
         HALOGAS  & $1.070 \pm 0.050$ & $1.110 \pm 0.110$  \\ 
        \hline   
         VIVA  & $1.190 \pm 0.170$ & $1.200 \pm 0.210$  \\ 
         %Note: M$_{dyn}$ $> 1.4\times10^{10} M_{\odot}$ (derived from the lower bound of $\cos (\theta_{inc})$)
    \end{tabular} 
\end{table}

%\begin{table*}{h}
%\centering
%\caption{Comparision of molecular gas mass, stellar mass, and dynamical mass}
%\label{table2}
    %\begin{tabular}{ccc}
        %\hline
         %Molecular gas mass  & Stellar mass  & Dynamical mass  \\
         %($M_{\odot}$)       & ($M_{\odot}$) & %($M_{\odot}$)   \\
         %\hline
         %$(2.02\pm0.18)\times10^{9}$ & $(1.32\pm0.51)\times10^{10}$ & $6.04\times10^{10}$\\
        %\hline   
         %Note: M$_{dyn}$ $> 1.4\times10^{10} M_{\odot}$ (derived from the lower bound of $\cos (\theta_{inc})$)
    %\end{tabular} 
%\end{table*}

\section{Discussion and conclusion}
\label{discussion}
%\subsection{Luminosity function and energy function}
To understand the gas kinematics in repeating and non-repeating FRB hosts, we compare our result of non-repeating FRB host with the previous study of HI detection \citep{Kaur2022, Michalowski2021} of repeating FRB host galaxies. However, we note that the HI and H$_{\rm 2}$ gas do not necessarily trace each other or are not necessarily physically associated. Therefore, the following comparison between the HI kinematics of the repeating FRB host galaxy and molecular gas kinematics is not strictly direct.
 The host galaxy of repeating FRB 20180916B \citep{Kaur2022} shows a distorted and merging kinetic structure of HI gas which is experiencing a minor merger. The merging gas increases the HI mass and leads to the burst of star formation in the outskirt of the host galaxy, suggesting a possible link between the FRB progenitor and the enhanced star formation \citep{Kaur2022}.

In \citet{Michalowski2021}, the asymmetric HI spectra of repeating FRB host galaxies i.e., NGC 3252 (the host of FRB 181030A) and M81 (the host of FRB 200120E), show the disturbed HI gas structures,  resulting in the recent enhancement of SFR due to the interaction between galaxies. However, the environment of FRB 200120E is puzzling because the FRB is localized in a globular cluster in M81 \citep{Kirsten2022}, which is an old system in contrast to the enhanced SFR due to the gas interaction.

In our work, the asymmetric CO spectrum (figure \ref{fig.1} right), the multiple spatial components in the velocity integrated intensity map (figure \ref{fig.2}), and the asymmetric velocity gradient in velocity structure (figure \ref{fig.4}) all indicate the disturbed kinematic structure of molecular gas in the host galaxy.
A comparison to the previous works \citep{Kaur2022, Michalowski2021} suggests that the disturbed kinetic gas structure could commonly appear in both repeating and non-repeating FRB, which suggests a possible link between the gas kinematics and the FRB progenitors.

Previous optical studies found that FRB host galaxies show diverse properties in the colour-magnitude diagram and BPT diagram \citep{Heintz2020, Bhandari2022}, implying no conclusive answer to the galaxy population of FRB hosts. Our study reveals a new perspective that disturbed gas kinematics are found in both repeating and non-repeating FRB host galaxies and suggests the possible link to their progenitors.
However, the observation of molecular gas in FRB host galaxies is still limited. Therefore, more observations are essential to understand the relationship between the gas kinematics and FRB progenitor.

%In our work, according to the gas kinematics, we conclude the molecular gas is undergoing smooth rotation, which aligns with the host galaxy of FRB180924B, an early-type spiral galaxy (\citep{Bannister2019}) with a relatively small SFR. Comparing the gas kinetic of two FRB host galaxies, we could imply the diverse progenitor and physical environment of different types of FRB host galaxies or different populations of progenitors.

%\section{Conclusion}
%\label{conclusion}
\section*{Acknowledgements}
We are very grateful to the anonymous referee for insightful comments.
We thank the ALMA staff for data acquisition and Casa technical manipulation.
T-YH acknowledges the University Consortium of ALMA-Taiwan Center for ALMA Science Advancement through a grant 110-2112-M-007-034-.
TH acknowledges the support of the National Science and Technology Council of Taiwan through grants 110-2112-M-005-013-MY3, 110-2112-M-007-034-, and 111-2123-M-001-008-.
TG acknowledges the support of the National Science and Technology Council of Taiwan through grants 108-2628-M-007-004-MY3 and 111-2123-M-001-008-.
BH is supported by JSPS KAKENHI Grant Number 19K03925.
This paper makes use of the following ALMA data: ADS/JAO.ALMA\#2019.1.01450.S. ALMA is a partnership of ESO (representing its member states), NSF (USA) and NINS (Japan), together with NRC (Canada), MOST and ASIAA (Taiwan), and KASI (Republic of Korea), in cooperation with the Republic of Chile. The Joint ALMA Observatory is operated by ESO, AUI/NRAO, and NAOJ.

%%%%%%%%%%%%%%%%%%%%%%%%%%%%%%%%%%%%%%%%%%%%%%%%%%
\section*{Data availability}
The data underlying this article can be obtained through ALMA Science Archive (\url{https://almascience.nrao.edu/aq/}).
%The parameters derived in this work are available in the article's online supplementary material.


\bibliographystyle{mnras}
\bibliography{ALMA} % if your bibtex file is called example.bib

%\begin{thebibliography}{99}
%\end{thebibliography}

%%%%%%%%%%%%%%%%%%%%%%%%%%%%%%%%%%%%%%%%%%%%%%%%%%

%%%%%%%%%%%%%%%%% APPENDICES %%%%%%%%%%%%%%%%%%%%%

% Don't change these lines
\bsp	% typesetting comment
\label{lastpage}
\end{document}